\documentclass[prb,twocolumn,amsmath,amssymb,floatfix]{revtex4}
\usepackage{graphicx}% Include figure files

\newcommand{\be}{\begin{eqnarray}}
\newcommand{\ee}{\end{eqnarray}}
\newcommand{\bra}[1]{\langle #1|}
\newcommand{\ket}[1]{|#1 \rangle}

\newcommand{\bfS}{{\bf S}}
\newcommand{\bfI}{{\bf I}}

\newcommand{\wbe}{\begin{widetext}}
\newcommand{\wee}{\end{widetext}}
\newcommand{\oncite}{\onlinecite}

\begin{document}
\draft

\title{Universal dynamics of quantum spin decoherence in a 
nuclear spin bath}

\author{Yi-Ya Tian, Pochung Chen and Daw-Wei Wang}

\address{Physics Department, National Tsing-Hua University, Hsinchu,
Taiwan, ROC }

\date{\today}

\begin{abstract}
We systematically investigate the universal spin decoherence dynamics of
a localized electron in an arbitrary nuclear spin bath, 
which can be even far away from 
equilibrium due to the weak nuclear-lattice interaction. We 
show that the electron spin relaxation dynamics
(as well as spin pure dephasing and Hahn echo decay)
can {\it always} have a universal behavior 
as long as the initial state is
composed of a sufficiently large amount of spin eigenstates.
For a given system, the pattern of the universal dynamics
depends on the complicated initial condition only via 
a {\it single} parameter,
which measures the amount of phase coherence
between different spin eigenstates in the initial state.
Our results apply even when
the number of the involved nuclei is not large, and therefore
provide a solid foundation in the comparison of theoretical/numerical 
results with the experimental measurement.
As an example, we also show numerical results for 
systems of noninteracting spin bath in zero magnetic field regime,
and discuss the features of universal decoherent dynamics.
\end{abstract}

%\pacs{PACS numbers:}
\maketitle
%%%%%%%%%%%%%%%%%%%%%%%%%%%%
%%%%%%%%%%%%%%%%%%%%%%%%%%%%
\section{ Introduction}
Localized spins in solid-state systems are one of the most
promising candidates for realizing the quantum computation
due to its long coherence time [\oncite{long T}]
and the possible scalability [\oncite{loss98}]. 
Recently, quantum control of single localized spin becomes experimentally 
feasible [\oncite{control}], 
but the nuclear spin bath induced decoherence, which is the 
dominant decoherent mechanism in the low temperature regime, 
is still hindering the further developments.
To study the effects of nuclear spin bath,
both analytical approaches [\oncite{theory,QSA,EOM}] and
numerical simulations [\oncite{ED,P-rep}] are developed
for different parameter regime.
The effect of dipolar interaction between nuclear spins 
are also studied in Refs. [\oncite{stoch}]. 
However, to the best of our knowledge, 
an uncontested conclusion about the spin decoherence
dynamics and its relation to the experimental measurement
is still unavailable, even though the deleterious effects of nuclear spin
have been verified in recent experiments [\oncite{exp,lukin}].

%%%%%%%%%%%%%%%%%%%%%%%%%%%%%%%%%%%%%%%
%%% Introduction and claims%%%%%%%%%%%%
%%%%%%%%%%%%%%%%%%%%%%%%%%%%%%%%%%%%%%%
From experimental side, the most crucial limitation results
from the fact that the initial nuclear spin configuration is
very little known nor controllable. This is a highly nontrivial problem, 
because even if a
thermalized spin bath is assumed in the beginning (as done
in most theoretical work [\oncite{QSA,EOM,stoch}]), 
any quantum measurement or manipulation
of electron spin can just destroy the equilibrium and
lead to a highly non-equilibrium nuclear spin 
dynamics. The coherent time of nuclear spin bath is known to be
extremely long (can exceed 1s in GaAs quantum well [\oncite{NMemory}] 
and 25s in  $^{29}$Si isotope [\oncite{ladd}]) 
and therefore it is very questionable if the nuclear spin bath could be 
well-thermalized for the next quantum 
measurement/manipulation  in a short time during the quantum computation 
process.
In order to have a meaningful comparison between theoretical results 
and the experimental 
measurement, the first and the most important question one should ask 
is if there could be any universal dynamics in such a system, which is 
insensitive to the details of initial nuclear spin configuration.

From theoretical side, answering above question is also 
very difficult because
the spin dynamics of one configuration can be very different from 
the other [\oncite{ED}] even though their initial configurations
are similar. Moreover, in a typical 
quantum dot system, the number of nuclei can be very huge 
($N\sim 10^{3-5}$), and hence it is also a significant challenge  
for ordinary numerical simulation to explore such huge phase space. 
These challenges are fundamentally
important to the understanding of the spin decoherence mechanism 
and to its future application in quantum computation.
However, to the best of our knowledge, there has no systematically
study in the literature to this important issue.

In this paper we address this issue 
by rigorously prove the existence of a generic and universal
electron spin decoherent dynamics in an arbitrary nuclear spin bath.
By ``universal dynamics'' we mean an electron spin evolution
which is of zero standard deviation over different initial condition
in the whole phase space. More precisely, we show that (1) the universality
of spin decoherence {\it always} exists if only the initial state
is composed of sufficient large amount of spin eigenstates, and
(2) for a given system, such universal dynamics depends
on the initial configuration only through a {\it single} parameter,
which measures the amount of phase coherence between spin 
eigenstates of the initial wavefunction. (3) The universality is ensured
by the large amount of phase space rather than the large value of nuclear
number, $N$, and therefore numerical simulation for a small size system
(say $N\sim 10-20$) can be still good enough to compare with a realistic
system of much more nuclei [\oncite{SmallN}]. Finally, (4)
the universality of spin dynamics applies to the decoherence of the diagonal 
part ($S_z$) as well as the off-diagonal part ($S_x$) of electron spin,
no matter in a free induction decay (FID) or in a Hahn echo decay.
Therefore our results resolves the fundamental problems in the comparison
of a theoretical calculation and an experimental measurement, and 
provide a new direction for the future study of the spin 
decoherence.
We also study the spin dynamics for systems of different electron/nuclear
spins, and find that the spin dynamics is mainly determined by the geometric
structure of the system density of states and is therefore insensitive to
the magnitude of nuclear spin.

This paper is  organized as fellows: In section II we describe the system 
Hamiltonian 
and the initial wavefunctions in our study.  In section III we show the 
universal dynamics of electron spin
relaxation. We study the universal dynamics by using both numerical and 
analytical methods. 
In section IV we discuss the microscopic origin of the universality. 
In section V, we generalize our consideration
to other spin systems. We conclude in section VI.

%%%%%%%%%%%%%%%%%%%%%%%%%%%%%%%%%%%%%%%%%%%%%%%%%%%%%%%%%%%%%%%%
%%%%%%%%%%%%%%%%%%%%%%%%%%%%%%%%%%%%%%%%%%%%%%%%%%%%%%%%%%%%%%%%

\section{Spin eigenstates and phase space}
\begin{figure}
\includegraphics[width=6cm]{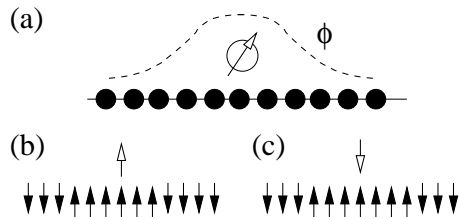}
\caption{
(a) Schematic pictures for electron spin (white circle) coupled to
a nuclear spin bath (black dots). The electron is assumed to be 
described by the orbital
envelope wave function $\phi$, and interacts with the nuclear spins 
(located at $\vec{r}_i$)
via a hyperfine interaction $A_i=A_0|\phi(\vec{r}_i)|^2$ where 
$A_0$ is the coupling strength.
(b) and (c) are two spin eigenstates with maximum/minimum
energies in zero magnetic field and $J_z=0$ case.}
\label{system}
\end{figure}

A general spin decoherence due to nuclear spin bath is described by the 
following Hamiltonian:
\begin{eqnarray}
\hat{H}&=&\hat{\bfS}\cdot\sum_i^N A_i\hat{\bfI}_i+
\hat{H}_{n-n}+\hat{H}_{Z},
\label{H}
\end{eqnarray}
where $\hat{\bfS}$ and $\hat{\bfI}_i$ are
respectively the dimensionless spin operators ($\hbar\equiv 1$)
of the localized electron and the nucleus at lattice site $i$.
$A_i$ is the hyperfine coupling strength, depending on the
wavefunction profile of the localized electron (Fig. \ref{system}(a)),
and we use  $A_i=A_0e^{-(3i/N)^2}$ for
our numerical calculation with $N$ being the number of nuclei.
{\bf We note that changing $N$ will not change the shape of the electron
wavefunction, and therefore increasing number of nuclear just reduce
the standard deviation of the spin dynamics instead of its average value.}
$\hat{H}_{n-n}$ and $\hat{H}_{Z}$ are respectively the 
interaction between nuclear spin and the Zeeman term
due to external magnetic field.
Even in the simplest case, where both $\hat{H}_Z$ and $\hat{H}_{n-n}$
are zero or neglected, 
the resulting dynamics due to the electron-nuclear coupling only
is still quite complex, because it
involves a huge amount of eigenstates in the Hilbert space.
In order to have a meaningful comparison between theoretical(numerical) 
results and the experimental observation, the first question 
one should ask is if there
could be any universal dynamics in such spin system, which is insensitive
to a general initial condition of the system and therefore can be
observed and repeatable in a realistic experiment. After all, 
it is very difficult to 
control and/or manipulate the spin configuration of nuclei in solid state
systems. This question, to the best of our knowledge, has
not been answered or even not addressed yet in the literature.
%--------
%------

It is convenient to use spin eigenstate,
$|\bfS\rangle_e\otimes|j\rangle_n\equiv |\bfS\rangle_e\otimes
|\{I_{1,z},I_{2,z},\cdots, I_{N,z}\}_j\rangle_n$, as the basis of
calculation, where $|\bf S\rangle_e$ is electron spin eigenstate
along certain direction (will be specified below) and
$\hat{I}_{i,z}$ is the nuclear spin eigenvalue along the direction
of magnetic field ($\hat{z}$) at the $i$th site.
%---
%Due to the conservation of total angular
%momentum, $J_z=S_z+\sum_{i=1}^N I_{i,z}$, we can consider a smaller subspace
%($\Omega$) with $J_z=0$ for the calculation throughout this paper.
%All results can be easily extended to systems of finite total angular moment.
%In this paper we always assume that the initial state 
%$|\psi_0\rangle$ is an eigenstate of $\hat{S}_z$ with largest eigenvalue $S$.
%---
For simplicity, in this paper, 
we assume the electron spin is initially polarized
only along $z$ or $x$ axis, and therefore 
a general initial wavefunction can be written to be: 
$|\psi_0\rangle_{x,z}=|S_{x,z}\rangle_e\otimes
\sum_{j=1}^{M_\Omega}a_j |j\rangle_n$,
where $a_j=r_je^{i\varphi_j}$ is the coefficient of the $j$th spin eigenstate
with phase $\varphi_j$ and amplitude $r_j$. Here
$|S_z\rangle_e\equiv|+\rangle_e$ and 
$|S_x\rangle_e\equiv\frac{1}{\sqrt{2}}\left(
|+\rangle_e+|-\rangle_e\right)$, and $M_\Omega$
is the size of the Hilbert space of nuclear bath.
%---
%an even smaller space ($\Gamma$), where all the spin
%eigenstates have $S_z=S$. As time evolves,
%the wavefunction goes out of the space $\Gamma$ and stay in a much larger
%space $\Omega$ with little probability to return, leading
%to a time-irreversable spin decoherent dynamics.
%-----
Using above expression, we consider the following three spin dynamics,
which are related to the spin relaxation, 
spin pure dephasing, and Hahn echo
decay respectively. The first two can be expressed as
$\langle S_{z,x}(t)\rangle
\equiv{}_{z,x}\bra{\psi_0}\hat{S}_{z,x}(t)\ket{\psi_0}_{z,x}
=\sum_{j=1}^{M_\Omega}r_j^2S_{j,j}^{z,x}(t)+\sum_{j\neq l}^{M_\Omega}r_{j}r_{l}
e^{-i(\varphi_{j}-\varphi_{l})}S_{j,l}^{z,x}(t)$, 
where $S_{j,l}^{z,x}(t)\equiv {}_n\langle j|\otimes
{}_e\langle S_{z,x}|\hat{S}_{z,x}(t)|S_{z,x}\rangle_e
\otimes|l\rangle_n$ is the matrix element.
$\langle S_z(t)\rangle$ can be very different from $\langle S_x(t)\rangle$ if the
nuclear spin is polarized by external magnetic field or with finite total angular 
momentum in a certain direction.
Similarly the Hahn echo decay is given by
$\rho_{+-}^H(\tau)\equiv {}_e\langle +|\hat{\rho}_H(\tau)|-\rangle_e$, where
the Hahn echo density matrix $\hat{\rho}_H(\tau)\equiv
{\rm Tr}_n\{U(\tau)|\psi_0\rangle_x{}_x \langle\psi_0|U(\tau)^\dagger\}
=\sum_j{}_n\langle j|U(\tau)|\psi_0\rangle_x{}_x\langle\psi_0|U(\tau)^\dagger 
|j\rangle_n$ and $U(\tau)\equiv e^{-iH\tau}\sigma_x e^{-iH\tau}$ 
[\oncite{Hahn,stoch}]. The characteristic time scale $T_2$ 
of pure dephasing is related to the single spin FID, while 
Hahn echo decay[\oncite{Hahn}] is 
usually used to extract single spin
behavior from an ensemble measurement.

%%%%%%%%%%%%%%%%%%%%%%%%%%%%%%%%%%%%%%%%%%%%%%%%%%%%%%%%%%%%%%%%
%%%%%%%%%%%%%%%%%%%%%%%%%%%%%%%%%%%%%%%%%%%%%%%%%%%%%%%%%%%%%%%%

\section{Universal dynamics}

In this section  we show that the universality
of spin decoherence always exists if only the initial state
is composed of sufficient large amount of spin eigenstates, and
for a given system, such universal dynamics depends
on the initial configuration only through a single parameter,
which measures the amount of phase coherence between spin 
eigenstates of the initial wavefunction. We first show the numerical results  
for spin relaxation, spin pure dephasing, and Hahn echo
decay respectively.  Then we rigorously give the proof of the universality. 

\subsection{Numerical study}
In order to explore the spin dynamics from different initial conditions in
the whole phase space, in this paper 
we allow both the amplitude, $\{r_j\}$, and the
phase, $\{\varphi_j\}$, to be independent variables and randomly chosen 
according to distribution functions
${\cal P}_r(r_j)$ and ${\cal P}_{\varphi}(\varphi_j)$ respectively.
The ensemble-averaged spin dynamics for $\langle S_z(t)\rangle$
becomes
$[\langle {S}_z(t)\rangle]\equiv\frac{[\langle S_z(t)\rangle]_{r,\varphi}}
{[\langle S_z(0)\rangle]_{r,\varphi}}$,
where $[f(r)]_{r}\equiv \int_0^1{\cal P}_r(r)f(r)dr$
denotes the average of a function $f(r)$, and similarly
$[f(\varphi)]_{\varphi}\equiv \int_0^{2\pi}{\cal P}_\varphi(\varphi)
f(\varphi)d\varphi$. $[\langle S_z(0)\rangle]_{r,\varphi}$ in the denominator
is for normalization. At the same time, the associated normalized 
standard deviation (NSD) is defined as follows:
\be
\sigma(t)&\equiv&\sqrt{\frac{[\langle \hat{S}_z(t)\rangle^2]_{r,\varphi}
-[\langle \hat{S}_z(t)\rangle]_{r,\varphi}^2}
{[\langle S_z(0)\rangle]_{r,\varphi}^2}}.
\label{sigma}
\ee
Similar definition of averaged dynamics as well as the NSD 
for $\langle S_x(t)\rangle$ and $\rho_{+-}^H(t)$ can be obtained easily. 
We note that, if the NSD of the averaged spin dynamics goes to zero
in the limit of infinite phase space, the averaged dynamics is also 
``the most probable'' dynamics with almost zero probability in the other
time-evolution behavior. As a result, we can define it as a universal
dynamics of the given system, independent 
(in the probability sense) of the details of
initial nuclear spin configuration. On the other hand, the system has
no universal dynamics if the NSD is of the order of one, since the average
value could not represent the characteristic dynamics of a general 
initial condition.

Before analytically studying the universality of spin dynamics in a general
system, it is more instructive to show some numerical results 
of the simplest system without magnetic field and nuclear spin interaction 
($\hat{H}_Z=\hat{H}_{n-n}=0$). 
We will first present the result for $\langle S_z(t)\rangle$ 
then the results for $\langle S_x(t)\rangle$ and $\rho_{+-}^H(t)$.
For the convenience of later discussion, 
we restrict the calculation inside a subspace, $\Gamma$, where the
total angular momentum, $J_z=S_z+\sum_{i=1}^N I_{i,z}$, is zero, and
choose ${\cal P}_r(r)=\gamma+(1-\gamma)\delta(r)$ and
${\cal P}_\varphi(\varphi)=(1-\xi)/2\pi+\xi\delta(\varphi)$. Here
$\gamma\in [0,1]$ can be understood as the probability to have 
a nonzero contribution in the subspace, $\Gamma$, 
while $\xi\in[0,1]$ is the probability
to have a phase coherence at a given value (set to be zero).
They satisfys the normalization condition: $\int_0^1{\cal P}_r(r)dr=
\int_0^{2\pi}P_\varphi(\varphi)d\varphi=1$ for all $\xi$ and $\gamma$. 

%------
\begin{figure}
\includegraphics[width=8.7cm]{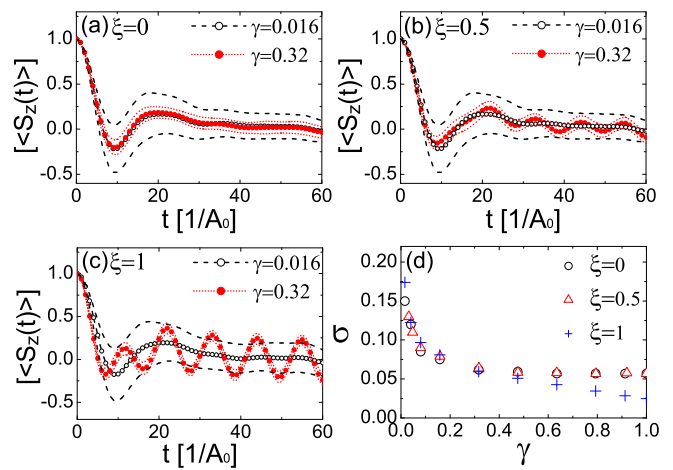}
\caption{(Color online)
(a)-(c) Averaged electron spin relaxation 
(symboled lines) for $\xi$=0, 0.5, and 1. 
respectively. Dashed and dotted lines are the uncertain range 
($[\langle S(t)\rangle]\pm\sigma(t)$)  
for $\gamma=0.016$ and 0.32 respectively.
(d) Time averaged NSD ($\sigma\equiv\lim_{T\to\infty}
\int_0^T\sigma(t)dt/T$) v.s. $\gamma$. $N=11$ in all figures.}
\label{universal}
\end{figure}
%------

In Fig.~\ref{universal}(a)-(c) we show the averaged electron spin
($S=\frac{1}{2}$) relaxation $[\langle S_z(t)\rangle]$,
in a noninteracting spin bath ($I=\frac{1}{2}$) 
with zero magnetic field for $\xi=0$, 0.5, and 1 with two 
different values of $\gamma$.
We observe that when $\gamma$ is small 
($\gamma=0.016$), meaning only a few spin 
eigenstates are involved in $|\psi_0\rangle_z$, 
the NSDs are very large, i.e., no universal dynamics.
This explains why in the literature different initial states can
result in very different time-evolution patterns [\oncite{ED}]. 
When $\gamma$ becomes larger ($\gamma=0.32$), the NSD
decreases in all figures (see also Fig.~\ref{universal}(d)), 
while the averaged dynamics 
also begin to show different behavior for different value of $\xi$: 
A two-decay curves for $\xi\to 0$ and a single mode oscillation 
for $\xi\to 1$. In fact, as we will show later, the NSD always decrease
to zero even $\gamma$ is finite, as long as the size of phase 
space becomes large enough.
These results indicate that
a universal dynamics can always be expected 
if the initial state is composed of 
a sufficiently large portion of spin eigenstates in the phase space, simply
due to the strong quantum interference effects. 
We could also show that the two decay time scale of Fig. 2(a) is due 
the the structure of the system density of states, and wil discuss that
in more details in the latter section.

\begin{figure}
\includegraphics[width=8.7cm]{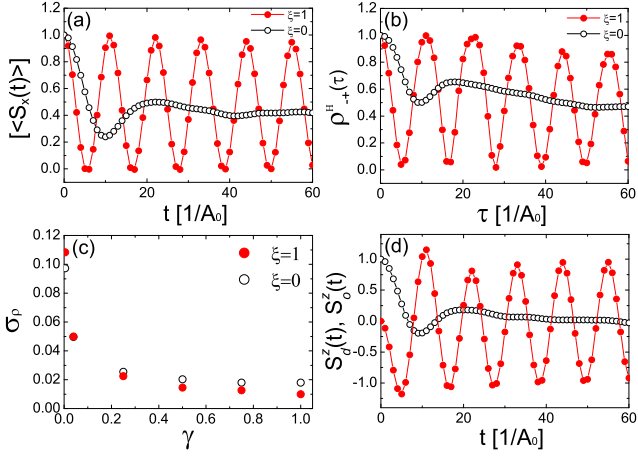}
\caption{(Color online)
(a) $[\langle S_x(t)\rangle]$ and (b) $\rho_{+-}^H(\tau)$
as a function of time $t$ and $\tau$ respectively (see text).
Lines with filled symbols and open symbols are for $\xi=1$ and $\xi=0$.
(c) The NSD at $\tau=10$ for Hahn echo decay.
(d) $S^z_d(t)$ (open circle) and $S^z_o(t)$ (filled circle) for the 
spin relaxation dynamics in Eq. (\ref{S_ave}).
}
\label{echo}
\end{figure}

In Fig.~\ref{echo}(a) and (b) we respectively plot pure dephasing 
($\langle S_x(t)\rangle$) and Hahn echo decay ($\rho_{+-}^H(\tau)$)
as a function of time for fully coherent ($\xi=1$, filled
circle) and fully incoherent ($\xi=0$, open circle) 
choice of initial condition.
To simplify the numerical calculation, we choose an initial wavefunction in a 
subspace of $\sum_{i=1}^N I_{I, z}=0$ for the pure dephasing ($\langle S_x(t)\rangle$, 
(a) ), and $J_z=0$ (same as $\langle S_z(t)\rangle$) for the Hahn echo decay ($\rho_{+-}^H(\tau)$, (b)). 
The electron for the former case is initially polarized in the $x$ direction so that the average total angular
momentum in $z$ direction, $\langle S_z(t)\rangle$, is still zero. As a result, 
the dynamics of dephasing, $\langle S_x(t)\rangle$, is different from the relaxation, 
$\langle S_z(t)\rangle$ due to the different choice of subspace where the initial wavefunction is defined.
We believe such convection is justified and will not affect any of our conclusion, 
because here we just use this numerical results as an example to understand the general properties 
of the universal dynamics. 
Full numerical results for any realistic situation will need a much larger phase space and much longer time.
Within this subspace, different initial wavefunctions still result in different dynamics ( not show here ). 
However, when the initial wavefunction is composed of sufficient large amount of eigenstates in the subspace, 
we again find a universal dynamics with almost zero NSD. In Fig 3(a) and (b), we show results for $\gamma=1$ 
for pure dephasing and Hahn echo decay in the two subspace described above. 
In (c), the NSD of the Hahn echo decay is plotted as function of $\gamma$ at $\tau=10$.
From these results, we find similar single mode oscillation for $\xi=1$, while a two-decay curves for $\xi=0$
in all the three dynamics 
$( \langle S_z(t)\rangle, \langle S_x(t)\rangle$ and $\rho_{+-}^H(\tau))$.
  
%--------
%---------------

%%%%%%%%%%%%%%%%%%%%%%%%%
\subsection{ Analytical study}
To analytically study the universal spin dynamics, we have to
do the ensemble-average first so that
\be
[\langle {S}_z(t)\rangle]
=S_d(t)+\frac{[r]_r^2}{[r^2]_r}
\left|[e^{-i\varphi}]_\varphi\right|^2 S_o(t)
\label{S_ave}
\ee
where we have used $[\langle S_z(0)\rangle]_{r,\varphi}=[r^2]_r\sum_j
\langle j|\hat{S}_z|j\rangle=[r^2]_r M_\Gamma$ in the normalization;
$S_d(t)\equiv M_\Gamma^{-1}\sum_jS_{j,j}(t)$ and
$S_{o}(t)\equiv M_\Gamma^{-1}\sum_{j_1\neq j_2}S_{j_1,j_2}(t)$
are the diagonal and off-diagonal maxtrix element of electron spin. 
We note that Eq. (\ref{S_ave}) indicates that the averaged spin dynamics
depends on the initial condition only via a {\it single}
parameter, $\beta\equiv \left([r]_r^2/[r^2]_r\right)
\left|[e^{-i\varphi}]_\varphi\right|^2$, which 
depends on the phase distribution function, ${\cal P}_\varphi$, much
more significantly than on the amplitude distribution function,
${\cal P}_r$, since $[r]_r^2\sim [r^2]_r$ for the usual function of
${\cal P}_r$ and $r\geq 0$.
We note that although the experimental preparation of a coherent
nuclear spin bath (i.e. finite value of $\beta$) is not easy at
the present stage, it has been realized how to control the coherent 
electron spin dynamics via interaction with a {\it single} 
nuclear spin in diamond [\oncite{lukin}]. Therefore, at least in a small 
quantum dot system, a coherent 
preparation and control of a few nuclear spin can be still realized.
In Fig. \ref{echo}(d) we show the time-evolution of both $S_d(t)$
and $S_o(t)$ of spin matrix element. Not surprisingly, 
they are of very different properties:
the diagonal part ($S_d(t)$) shows a clear two-decay process: with
a fast decay in short time and a slow decay in long time. However,
the off-diagonal part ($S_o(t)$) does not decay at all, and shows
a single mode oscillation. It is easy to see that
the numerical results shown in Fig. \ref{universal}(a)-(c) can be
obtained as a superposition of $S_d(t)$ and $S_o(t)$, just as suggested
by Eq. (\ref{S_ave}). Numerical comparison between these two approaches
(ensemble average before and after time-revolution) agree excellently
well (not shown here), showing that only a single parameter,
$\beta$, is necessary to reproduce all the ensemble averaged 
spin relaxation dynamics.

In order to exam if the ensemble-averaged results of Eq. (\ref{S_ave})
is a universal dynamics, we need to calculate
the fluctuation (NSD, Eq. (\ref{sigma})) about this average. 
For simplicity, we first study the case in the completely random phase limit,
say $[e^{i\varphi}]_\varphi=0$. We then have
$[\langle \hat{S}_z(t)\rangle]_{r,\varphi}^2=M_\Gamma^2
[r^2]^2S_d(t)^2$ according to Eq. (\ref{S_ave}).
After some algebra we can derive:
$[\langle \hat{S}_z(t)\rangle^2]_{r,\varphi}
-[\langle \hat{S}_z(t)\rangle]_{r,\varphi}^2
=([r^4]_r-2[r^2]_r^2)\sum_jS_{j,j}(t)^2+[r^2]_r^2
\sum_{j,l}S_{j,l}S_{l,j}$. Since we are interested in the
upper bound of the NSD, we may use
$\sum_{j,l}S_{j,l}(t)S_{l,j}(t)=\sum_j\langle j |\hat{S}_z(t)
\hat{P}_\Gamma\hat{S}_z(t)|j\rangle\leq\sum_j\langle j |\hat{S}_z(t)^2|j\rangle
\leq S(S+1)\sum_j\langle j |j\rangle=S(S+1)M_\Gamma$,
where $\hat{P}_\Gamma\equiv\sum_l|l\rangle\langle l|$ is
to project a state onto the subspace $\Gamma$ with electron spin eigenvalue
$S_z=S$. Here we have used the fact that for any state, the expectation value
of $\hat{S}_z(t)^2$ must be equal or smaller the expectation value
of total electron spin $\bfS^2$, which is, however, a conserved quantity
of our system (see Eq. (\ref{H})). Similarly we also have
$\sum_jS_{j,j}(t)^2\leq \sum_{i,j}|S_{i,j}(t)|^2\leq S(S+1)M_\Gamma$,
and therefore $[\langle \hat{S}_z(t)\rangle^2]_{r,\varphi}
-[\langle \hat{S}_z(t)\rangle]_{r,\varphi}^2\sim {\cal O}(M_\Gamma)$. In other
words, after devided by $[\langle S_z(0)\rangle]_{r,\varphi}^2
\propto M_\Gamma^2$, we find $\sigma(t)\propto M_\Gamma^{-1/2}$ and therefore
goes to zero in the limit of $N\gg 1$ or $M_\Gamma\to\infty$.

We can also apply similar method to study the NSD of a phase coherent
initial state, i.e. $[e^{i\varphi}]_\varphi\neq 0$. After some algebra,
the expansion of $[\langle \hat{S}_z(t)\rangle^2]_{r,\varphi}
-[\langle \hat{S}_z(t)\rangle]_{r,\varphi}^2$ will have two additional
summations (besides of the two shown above) with nonuniversal prefactors: 
first, we have $\sum_{i,j,l}S_{i,l}(t)S_{l,j}(t)=
\langle V|\hat{S}_z(t)\hat{P}_\Gamma\hat{S}_z(t)|V \rangle
\leq\langle V|\hat{S}_z(t)\hat{S}_z(t)|V\rangle
\leq S(S+1)\langle V|V\rangle
=S(S+1)\sum_{i,j}\langle i|j\rangle=S(S+1)M_\Gamma$,
where we define $|V\rangle\equiv\sum_j|j\rangle$ as an auxiliary state.
Secondly, $\sum_{l,j}S_{l,l}(t)S_{l,j}(t)\leq
\sqrt{\sum_lS_{l,l}(t)^2\sum_l|\sum_j S_{l,j}(t)|^2}
\leq M_\Gamma^{1/2}\sqrt{\langle V|\hat{S}_z(t)\hat{P}_\Gamma
\hat{S}_z(t)|V\rangle}
\leq M_\Gamma^{1/2}\sqrt{\langle V|\hat{S}_z(t)^2|V\rangle}
\leq \sqrt{S(S+1)}M_\Gamma$,
where we have used the fact that the inner product of two
vectors must be equal or smaller than the product of their length.
Therefore, after renormalized by the initial spin average,
we find $\sigma(t)\propto M_\Gamma^{-1}$ and becomes to zero in
large system size just as for the complete random phase case ($\xi=0$).
 From the above results, we conclude that no matter
how much phase coherence between spin eigenstates of the initial wavefunction,
the spin relaxation dynamics
can be always universal (with zero NSD) in the limit of infinitely large
phase space ($M_\Gamma\gg 1$). Similar derivation for dynamics
of pure dephasing ($\langle S_x(t)\rangle$) and Hahn echo decay
($\rho^H_{+-}(\tau)$) can be obtained straightforwardly.

\section{ Microscopic origin of universality}

%--------
\begin{figure}
\includegraphics[width=8.7cm]{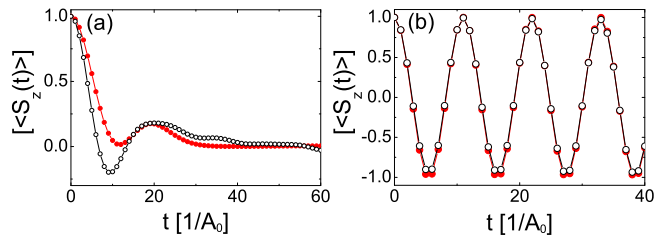}
\caption{(a) Comparison between the calculated $\langle S_z(t)\rangle$ 
(open circle) with $\xi=0$ and $\gamma=1$ and $\langle S_z(t)\rangle_{\rm DOS}$
(filled circle) obtained from density of states only.
(b) Comparison between the calculated $\langle S_z(t)\rangle$ with
$\xi=1$ and $\gamma=1$ and the result obtained by including 
$|E_{\rm max/min}\rangle$ only (see text).
}
\label{fit}
\end{figure}
%------

After concluding the universality of the most general spin relaxation
system (Eq. (\ref{H})), in the rest of this paper we return to 
a less general case, zero magnetic field
and noninteracting spin bath ($\hat{H}_Z=\hat{H}_{n-n}=0$), to study 
the microscopic origin of the universal spin relaxation curves shown 
in Fig. \ref{universal}(a)-(c).
We first rewrite $\langle \hat{S}_z(t)\rangle$
in terms of energy eigenstates, $|E\rangle$:
\be
\langle \hat{S}_s(t)\rangle&=&\int dE{\cal D}(E)\int dE'{\cal D}(E')
C_EC_{E'}^\ast
\nonumber\\
&&\times\langle E|\hat{S}_z|E'\rangle\,e^{-i(E-E')t},
\label{S_t_E}
\ee
where ${\cal D}(E)$ is the density of states (DOS) of the system and
$C_E\equiv\ \langle \psi_0|E\rangle$. According to our numerical calculation,
we observe that the matrix element, $\langle E|\hat{S}_z|E'\rangle$,
varies almost randomly for different energies. {\bf Since now
we are interested in the simplest possible explanation for the features
of electron spin dynamics, we can first neglect such structureless random
matrix element for simplcity. As we will see later, it turns out that
this simplication does bring a very useful understanding of the universal
spin dynamics.}

%--------
\begin{figure}
\includegraphics[width=8.5cm]{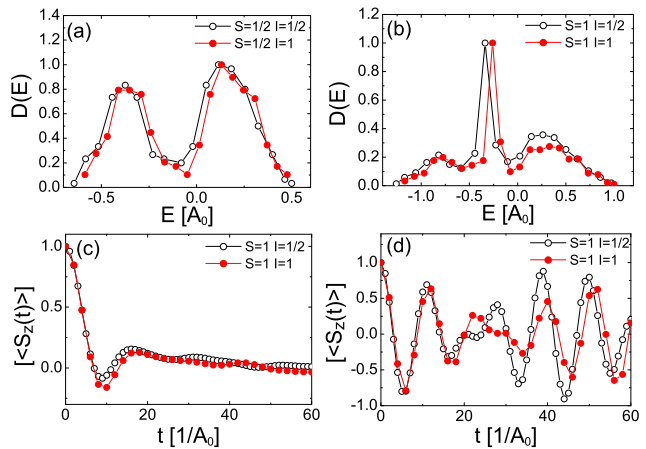}
\caption{(Color online)(a) and (b) are the density of states of systems with  
different electron/nuclear spins. Note that in both figures, 
the energy axis for $I=1$ cases (filled circle) 
have been rescaled by a factor $1/2$ in order
to fit the same scale as $I=1/2$ case. 
(c) and (d) are ensemble-averaged spin relaxation curves
for $S=1$ and $\gamma=1$ case, with $\xi=0$ and $\xi=1$ respectively.
For comparison, results of different nuclear spins are shown 
together after rescaling the time axis (see above).}
\label{spin}
\end{figure}
%---------
For the case when initial wavefunction $|\psi_0\rangle$,
is totally randomly distributed in the phase space, $\Gamma$, i.e.
$\gamma=1$ and $\xi=0$, we can further assume that
$\langle\psi_0|E\rangle$ is also independent of energy $E$ in above equation.
As a result, Eq. (\ref{S_t_E}) can be approximated by
$\langle {S}_z(t)\rangle_{\rm DOS}\equiv
\left|\int dE{\cal D}(E)e^{-iEt}\right|^2$, which is just power spectrum
the density of state.
In Fig. \ref{fit}(a) we show the full numerical result of
$\langle S_z(t)\rangle$ for $\xi=0$
compared to $\langle {S}_z(t)\rangle_{DOS}$ given above.
One can see that the later can qualitatively reproduce all the
important structure of the full numerical results. This agreement helps
us to conclude that the decay time of $\langle S_z(t)\rangle$
is mainly determined by the width of DOS peaks (see Fig. \ref{spin}(a)),
while the time scale of the second peak of $\langle S_z(t)\rangle$ is
given by the energy separation between the two peaks in DOS.

As for the single mode oscillation shown in Fig. \ref{universal}(c) for
full spin coherent initial state ($\xi=1$), we can apply similar study but
notice that the coefficient, $C_E$, is not a constant for all eigenstate
energy any longer. Our numerical calculation shows that $|C_E|$
is very small ($\propto M_\Gamma^{-1/2}$) for all energy except near
$E=E_{\rm max/min}$, where $|E_{\rm max/min}\rangle$
is the eigenstates of the top(maximum) and bottom(minimum) of the eigenenergy
band. This is because 
$|E_{\rm max/min}\rangle$ has a large overlap with some particular
spin eigenstate (as shown in Fig. \ref{system}(a) and (b)) and the
overlapping coefficients are not cancelled out due to the same
phase in a full spin coherent initial state ($\xi=1$). 
In Fig. \ref{fit}(b), we compare the numerical result of the universal
dynamics ($\gamma=1$) of a full coherent
initial state ($\xi=1$) and the result calculated by using 
$|E_{\rm max/min}\rangle$ only (with proper nomalization). We 
find the agreement
is excellent, predicting the same oscillation frequency and even the same
phase. {\bf The agreement justifies the approximations used in the derivations
after Eq. (4) and also shows that} the 
universal behavior of the spin relaxation
dynamics can be simply explained by the structure of density of states
and the two special spin configurations as shown in 
Fig. \ref{system}(a) and (b). As for results with $0<\xi<1$, it can be 
also explained well by a linear combination of above two results, as suggested
by Eq. (\ref{S_ave}).

%%%%%%%%%%%%
\section{Results for different spins}
After systematically investigating the spin relaxation dynamics for
an spin-half electron inside a spin-half nucear spin bath, here we
further extend the study of universality to systems of different 
electron/nuclear spins. 
In Fig. \ref{spin}(a) and (b), we show the density of states
for $(S,I)=(\frac{1}{2},\frac{1}{2})$, $(S,I)=(\frac{1}{2},1)$,
$(S,I)=(1,\frac{1}{2})$, and $(S,I)=(1,1)$ in different curves. For
the convenience of comparison, we rescale the energy scale in each plot
and normalize the hieght of DOS by the total size of phase space, $\Gamma$.
Surpisingly we find that the DOS structure is almost the same for
different nuclear spins $I$ as long as the electron spin $S$ is the same. 
This reflects the
fact that the total Hilbert space of the nuclear spin bath has been
large enough due to the number of nucluei so that
the spin degrees of freedom does play very little role in the structure
of energy spectrum. 
Analysing the energy eigenstate configuration,
we find the spin configuration near the degeneracy regime 
(position of the peaks) are related to if the the central nuclei
spin configuration is polarized and parallel (or anti-parallel)
to the electron spin. Similar observation also applys to 
the triple peak structure in Fig. \ref{spin}(b) for $S=1$: 
In Fig. \ref{spin}(c) and (d), we show the spin relaxation curves for
$S=1$ with spin phase random ($\xi=0$) and spin phase
coherent ($\xi=1$) initial wavefunctions respectively, after properly
rescaling the horizontal axis.
One can see that results in (c) are very similar to the
spin half case (Fig. \ref{universal}(a)), while it shows a beating
oscillation for a coherent initial wavefunction (d). The rescaled
time-evolution for $I=\frac{1}{2}$ and $I=1$ are very similar, except for
a small phase twist. We then conclude that the the spin relaxation
dynamics is insensitive to the nuclear spin degrees of freedom,
consistent with our earlier statement that the universal 
spin dynamics is independent of the nuclear spin configuration.
Our results for $S=I=1$ can be also applied to the study of spin dynamics in
the mixtures of spinful cold atoms in all-optical trap, where the
localized ``electron'' and the ``nucluei'' can be prepared easily by using
optical lattice with proper wavelength difference. The advantage for
cold atom system is that the initial spin configuration can be
prepared easily and the coupling strength, $A_0$, can be tuned via optical
Feshbach resonance and/or other method.

%%%%%%%%%%
\section{Conclusions}
In this paper we rigorously prove that the electron 
spin decoherence due to nuclear spin bath can be always 
universal if only coupled by sufficient large amount of spin eignestates.
There are several features about the universal dynamics 
that we want to emphasize:
First, in the derivation above, we do not rely on
any particular form of the distribution function
(${\cal P}_r$ and ${\cal P}_\varphi$),
hence the universality of spin dynamics 
is independent of the nuclear spin configuration.
However, if the initial state is composed of only finite numbers
of spin eigenstates (as done in the literature), 
our derivation will fail since $[r^2]_r\to 0$ 
in the denominator of $\sigma(t)$, i.e., no universal dynamics can be expected.
Secondly, the universality does not rely on any particular Hamiltonian,
so our conclusion also applies to systems which include nuclear spin
interaction, finite magnetic field, or any other more complicated system. 
Different system Hamiltonians just bring different
averaged results of spin dynamics, but the huge phase space
({\it not} necessary the huge nuclear number) can {\it always} ensure 
it to be the most probable one
regardless of the details in the initial condition. 
Such important results lead to another conclusion that
a numerical simulation of a much
smaller system (say $N\sim 10-20$) can still have large enough phase space
($M_\Omega=(2S+1)\times(2I+1)^N\sim10^{3\sim6}$) 
and hence gives similar results as given by macroscopic number of 
nuclei [\oncite{SmallN}].
{\bf Excellent agreement between our results of small size calculation 
(Fig. \ref{universal}(a) with $N=11$) 
and a meanfield type calculation of a much larger system (for example,
Fig. 4 of Ref. [\oncite{P-rep}] with $N=2000$) ensures the existence 
of such scale-indepedent universal dynsmics.}
Our results therefore make a realistic comparison between 
a theoretical calculation and experimental data possible, leaving only
a single unknown parameter, $\beta$, as a fitting parameter.
{\bf (For example, in the Fig. 4 of Ref. [\oncite{P-rep}], $\beta=0$
is expected due to the thermalized initial bath.)}
Finally, our derivation relies on the fact that electron spin is a conserved
quantity with an upper-bounded eigenvalue (not scaled with system size). 
This may explain why
spin eigenstate can be a special basis for studying universal physics
and restricts a naive application of our results to the 
relaxation dynamics of other physical quantities.

{\bf
It is also worthy to note that the universal dynamics may bear a 
close relationship with the quantum central limit theorem (QCLT).
QCLT has been used to study the quantum state estimation without 
using a large ensemble [\oncite{QCLT1}]
and to explain why quantum and classical random walks possess different 
behaviors [\oncite{QCLT2}].
It is nature to conjecture that the existence of the universal dynamics 
and the reason why a small size
system can already capture the behavior of the macroscopic system 
can be understood in the context 
of QCLT. For example, in Ref [\oncite{QCLT1}], it was 
pointed out that a quantum state estimation
with small error using small size ensemble is possible. 
This is clearly resemble to our work, where a 
small size system can capture the universal dynamics of a 
macroscopic system.
However, such a connection is not at all transparent in the details of
the approach. Further study about such an interesting connection
is beyound the scope of this paper but could be very interesting 
for furure investigation.}

We appreciate fruitful discussion with S. Das sarma, 
X. Hu, S. K. Saikin, L. J. Sham, and J. M. Taylor. 
This work is supported by NSC Taiwan.

%%%%%%%%%%%%%%%%%%%%%%%%%%%%%%%

\end{document}